\documentclass[aps,10pt,prl,twocolumn,letterpaper,superscriptaddress]{revtex4-2} 

\usepackage{graphicx} 
\usepackage{bm} 
\usepackage{amsmath}
\usepackage{xcolor}

\definecolor{PRLblue}{rgb}{0.18,0.18,0.57}
\usepackage[colorlinks=true,allcolors=PRLblue]{hyperref} 
\graphicspath{ {./}{figs/}}
\usepackage{fancyhdr}
\begin{document}
\title{Coactive-Staggered Feature in Weyl Materials for Enhancing the Anomalous Nernst Conductivity}


\author{Vsevolod Ivanov}
\affiliation{Molecular Foundry, Lawrence Berkeley National Laboratory, Berkeley, CA 94720, USA}
\affiliation{Accelerator Technology and Applied Physics Division, Lawrence Berkeley National Laboratory, Berkeley, CA 94720, USA}
\author{Ella Banyas}
\affiliation{Molecular Foundry, Lawrence Berkeley National Laboratory, Berkeley, CA 94720, USA}
\affiliation{Department of Physics, University of California, Berkeley, CA 94720, USA}
\author{Liang Z. Tan}
\affiliation{Molecular Foundry, Lawrence Berkeley National Laboratory, Berkeley, CA 94720, USA}

\begin{abstract}	
Power generation through the anomalous Nernst effect in topological Weyl materials has several advantages over conventional thermoelectrics due to the transverse geometry. However, the magnitude of the anomalous Nernst conductivity (ANC) in most known materials is too small to be of practical use, and there exist few guiding principles for finding materials with optimal thermoelectric properties. This work shows that the ANC is maximal when there is a ``coactive-staggered" feature in the anomalous Hall conductivity (AHC). It is shown that a minimal arrangement of two Weyl pairs leads to such a feature, and tuning the separations between the pairs controls the temperature at which the ANC is maximal. Several methods are proposed for creating such arrangements of Weyl points starting from Dirac semimetal materials. It is also demonstrated how an existing coactive-staggered AHC in a Heusler material can be exploited, by collectively tuning the positions of the Weyl points through strain to further enhance the ANC. A modest 20\% amplification of the ANC is achieved, even with relatively minor changes in Weyl point positions. 
\end{abstract}

\maketitle

\section{I. Introduction.}

Materials hosting topological Weyl points have recently reinvigorated the search for new thermoelectric materials due to their potential to exhibit large anomalous Nernst conductivities (ANC). In the Nernst effect, electric current is generated transverse to the temperature gradient, leading to a higher potential efficiency than standard thermoelectrics as well as a simpler configuration for device applications \cite{weyl-thermoelectrics}. Topological materials with increasingly larger ANC are being discovered, including an ANC of $\sim$ 4 A m$^{-1}$ K$^{-1}$ in Co$_2$MnGa \cite{giantANE-co2mnga},
10 A m$^{-1}$ K$^{-1}$ in YbMnBi$_2$ \cite{ybmnbi2}, 15 A m$^{-1}$ K$^{-1}$ in UCo$_{0.8}$Ru$_{0.2}$Al \cite{ucra-sciadv}, and candidates identified in high-throughput computational searches~\cite{samathrakis2021}. Despite the numerous material discoveries and considerable demand for more efficient thermoelectrics, overarching materials design principles for large ANC have not yet emerged.

It is now well-understood how individual Weyl points arising from a single pair of bands crossing will contribute to the Berry curvature (BC), and in turn AHC and ANC. However, there has been comparatively little analysis on the interactions of Weyl points arising from separate sets of bands, or the collective behavior of large numbers of Weyl points. In Weyl metals, numerous bands crossing the Fermi level give rise to many Weyl points which all shift relative to each other when the material is perturbed by doping, strain, or interfacing. Understanding the multi-band Weyl physics in these materials is essential for future efforts to predict useful topological Weyl thermoelectrics. 

Materials with large ANC tend to possess several essential properties, including strong spin-orbit coupling, flat bands, and large BC in the Brillouin zone (BZ) \cite{ucra-sciadv}, but it is not yet clear how to engineer such systems. In this work we demonstrate how the ANC in Weyl materials can be enhanced by a two-peak feature in the AHC, which we term ``coactive staggered" (CS). 
In Section II, we present a minimal $\bm{k} \cdot \bm{p}$ model with four Weyl points that can generate a CS feature in the AHC, and demonstrate enhancement in the ANC, showing that for each possible energy separation of the Weyl points there will be a different optimal temperature where the ANC is maximal. Section III discusses engineering such configurations of Weyl points in Dirac semimetals using readily available experimental techniques. Then in Section IV, we show how in a material where a large number of Weyl points contribute to a CS feature in the AHC, the ANC can be further enhanced by collectively perturbing the Weyl positions.
Finally in Section V we conclude by discussing how this kind of control over the Weyl positions can be used to tune thermoelectric materials for applications at specific temperatures, and comment on how this new design principle for materials with large ANC can guide future directions of study.

\section{II. $ k\cdot p$ Model.}

A fundamental aspect of topological features in the electronic structures of materials is their non-trivial BC, which affects electron motion in these materials. In magnetic topological Weyl semimetals, the inherent magnetism leads to an intrinsic anomalous Hall conductivity (AHC), while an additional thermal gradient gives rise to ANC. It can be shown that for a given temperature $T$, the anomalous Nernst coefficient $\alpha_{ij}(T,\mu)$ can be related to the zero-temperature anomalous Hall coefficient $\sigma_{ij}(0,\epsilon)$ as 
\begin{equation}
    \alpha_{ij}(T,\mu) = -\frac{1}{e} \int d\epsilon \left( \frac{\partial f_{\text{FD}} }{\partial \mu} \right) \sigma_{ij}(0,\epsilon) \frac{\epsilon-\mu}{T},
    \label{ane-int}
\end{equation}
where $f_{\text{FD}}$ is the Fermi-Dirac distribution, $e$ is the electron charge, and $\mu$ is the chemical potential.


In the low-temperature limit, this expression reduces to the well-known Mott relation, $\alpha_{ij}(\epsilon) = (\pi^2/3) (k_B^2 T/e) \sigma^\prime_{ij}(\epsilon)$, between the ANC and energy derivative of the AHC \cite{ane-berry}
where $k_B$ is Boltzmann's constant. At higher temperatures this no longer holds, but can still be used to qualitatively understand the relationship between the AHC and ANC of Weyl materials.

\begin{figure}[b]
    \includegraphics[width=1.0\columnwidth]{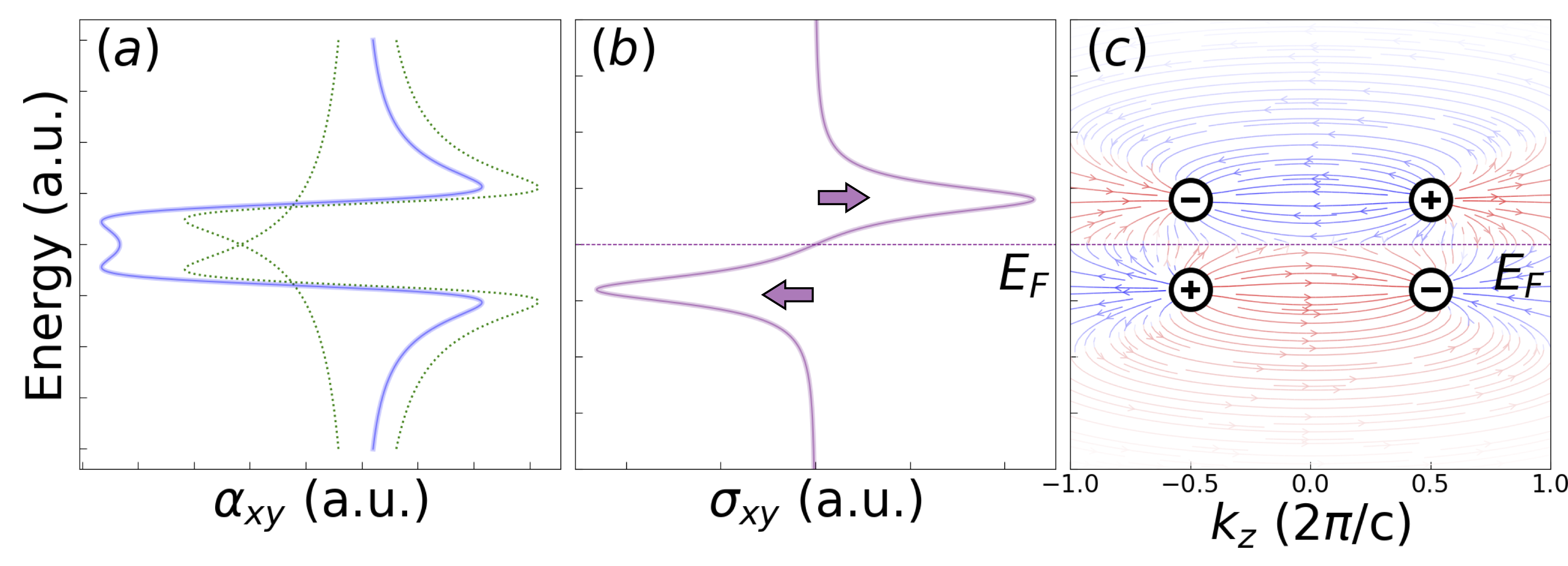}
    \caption{
    A minimal four-Weyl model (c), with positive (negative) BC flux shown in red (blue), with a CS feature in the AHC (b) as indicated by the arrows, leading to a peak in the ANC (a). Green lines in (a) show the separate contributions to the ANC above/below $E_F$ which combine to form the peak.
    }
    \label{cartoon}
\end{figure}

In a magnetic Weyl semimetal, the BC from a single pair of Weyl points at energy $E_\text{Weyl}$ generates a single peak in the AHC at that energy, whose width broadens with increasing temperature \cite{noky}.
Based on the Mott relationship 
, the large slope in the AHC with respect to energy will result in strong positive (negative) ANC just above (below) $E_\text{Weyl}$. Clearly a locally maximal ANC arises when the greatest possible change of the AHC occurs in a narrow energy window, e.g. from a large positive to a large negative peak \cite{supplement}. Figure \ref{cartoon}b shows a schematic of such an AHC around the the Fermi energy $E_F$, which we term ``coactive staggered" (CS), describing the energy separation between the two peaks. Each peak in the AHC results in two peaks in the ANC, with the inner peaks combining coactively at $E_F$ to yield an enhanced ANC (Fig. \ref{cartoon}a). 

In order for a material's AHC to have two opposite-sign peaks around $E_F$, it must have oppositely oriented fluxes of BC above and below $E_F$ (Fig. \ref{cartoon}c). This can arise from two (or more) sets of Weyl point pairs respectively located above and below $E_F$, such that the $k$-space separation of chiralities above $E_F$ is inverted with respect to the lower set. 
Throughout the paper we use the term CS to interchangeably refer to either the two-peak feature in the AHC or the arrangement of Weyl pairs that generates it, which can possibly involve a large number of Weyl points. 


This coactive aspect can be better understood by using an effective $k\cdot p$ model describing the linear dispersion in the vicinity of the Weyl nodes \cite{Zyuzin2016}. Each pair of Weyl points separated along the $k_z$-direction is described by the Hamiltonian
\begin{align}
H(\bm{k})_+ &= +\hbar C(k_z - Q) - \hbar v \bm{\sigma}\cdot (\bm{k} - Q \hat{k}_z)\nonumber\\
H(\bm{k})_- &= -\hbar C(k_z + Q) + \hbar v \bm{\sigma}\cdot (\bm{k} + Q \hat{k}_z),
\label{zyuzin-model}
\end{align}
where $2Q$ is the separation between the Weyl points, $v$ is the Fermi velocity, and $C$ controls the tilting, with $|C|<|v|$ describing Type-I and $|C|>|v|$ describing Type-II Weyl points.

In this model, the zero-temperature limit of the anomalous Hall contributions $\sigma^{\pm}_{xy}$ from the Weyl point of chirality $\pm$ within each pair can be written \cite{Zyuzin2016}
\begin{align}
&\sigma^{\pm}_{xy} = \mp \frac{e^2}{8\pi^2} \int_{\Lambda-Q}^{-\Lambda-Q}\Bigg[ \text{sign}(k_z) \Theta(v^2k_z^2-(\pm Ck_z-\mu)^2)\nonumber\\
&+ \frac{vk_z}{|\pm Ck_z-\mu|} (1-\Theta(v^2k_z^2 - (\pm Ck_z-\mu)^2) ) \Bigg] dk_z,
\label{type2sigma}
\end{align}
where $\Theta$ denotes the Heavside-theta function and $\Lambda$ is a finite momentum cutoff. The total AHC can be obtained by adding the contribution $\sigma_{xy} = \sigma^{+}_{xy} + \sigma^{-}_{xy}$ from each pair. From there the ANC is computed by evaluating the integral in Eq. \ref{ane-int}.

\begin{figure}[ht]
    \includegraphics[width=1.0\columnwidth]{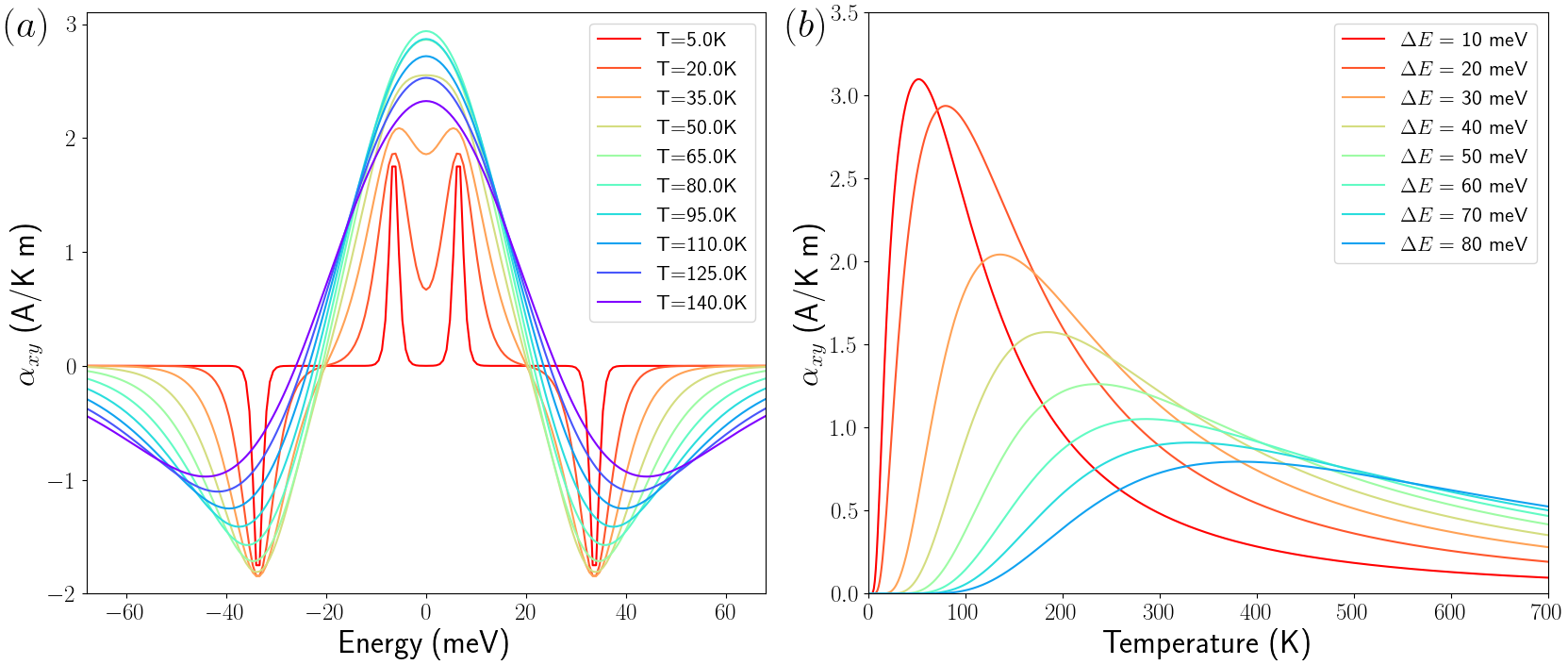}
    \caption{
        Plots of the (a) ANC vs energy at different temperatures for a model with fixed Weyl pair energy separation $\Delta E$, and (b) ANC vs temperature for different energy separation in the minimal CS model.}
    \label{ane}
\end{figure}

We consider the minimal CS configuration of four Weyl points (Fig. \ref{cartoon}c), separated in the BZ by $Q=0.1$ in units of $2\pi/c$ (c = 5.0 \r{A}), with the pairs located at $\Delta E = \pm 20$ meV around the Fermi energy. We use untilted Type-I Weyl points with $v = 0.05$ and $C = 0$ in units of Ry*$2\pi/c$, although the results do not depend significantly on the Weyl tilt. 

The resulting ANC for this model computed using Eq. \ref{ane-int} is shown in Figure \ref{ane}.
At low temperature, each Weyl pair creates two peaks above and below the Fermi energy (Fig \ref{ane}a), with the two positive peaks around $E_F$ being completely non-overlapping. With increasing temperature, these peaks broaden to combine at $E_F$ into a single peak that becomes maximal at $T=80$~K. For any given energy separation $\Delta E$ of the Weyl point pairs, there is a particular temperature for which the ANC is maximized, as shown in Figure \ref{ane}b. 
As $\Delta E$ is increased, the maximal possible value of ANC decreases, while the temperature at which the maximum occurs rises. 

We show that under general conditions, the AHC spectrum ($\sigma_{ij}(0,\epsilon)$) which maximizes the ANC response is in fact of the two-peaked CS shape (Supplemental Material). While this shape most naturally arises from the CS Weyl pair distribution discussed above, it can arise from other distributions of Berry curvature as well. For an AHC with an optimal CS shape, the precise upper bound on the ANC can be shown to be 
\begin{equation}
    \alpha_\text{max} = 0.815 \frac{k_B}{e} \bar{\sigma}  \Delta E_\text{opt}^{-\frac{1}{2}},
    \label{bound}
\end{equation}
where $\bar{\sigma}$ is the size of the zone-averaged anomalous Hall signal in the material. The peak separation in the optimal AHC scales linearly with temperature as $\Delta E_\text{opt} = 3.087 k_B T$.
This suggests that besides increasing the strength of Berry curvature sources ($\bar{\sigma}$), there is also the possibility of tuning the energy separation in CS Weyl point configurations in order to tune the ANC to a particular temperature and raise its maximal value, which will be explored in Section IV. We note here that $\Delta E$ for any given material should be evaluated in the context of the finite-temperature band structure (as should the ANC in general), as electronic band structures have implicit temperature dependence through the magnetization of the material.

\section{III. Engineering CS Weyls.}

While there are many materials which possess Weyl points and several methods have been developed for generating Weyl points \cite{weyl-create1}, having a CS AHC is a more stringent constraint. Pairs of Weyls at different energies must have oppositely oriented chiralities and arise in different bands, without any other pairs present that would cancel their BC contributions. Beyond this, there are of course other considerations requisite for sizeable topological transport properties, such as large momentum-space separation of the Weyl points and a large density of states at $E_F$. 

One approach to creating a CS Weyl configuration begins with a material that features a Dirac point, and uses symmetry-breaking to produce the desired Weyl points. 
Dirac semimetals are materials possessing both time-reversal ($\mathcal{T}$) and inversion ($\mathcal{I}$) symmetries, which host a four-fold degeneracy with linear dispersion called a Dirac point \cite{weyl-RMP,3D-DSM}. Excitations near this point can be described by the four-fold Hamiltonian of a Dirac fermion, which in turn can be decomposed into a pair of $2 \times 2$ Hamiltonians describing Weyl fermions \cite{volovik, savrasov-fermi-arc} of opposite chirality. If the $\mathcal{T}$ and/or $\mathcal{I}$ symmetry is broken, the Dirac point splits into equal numbers of positive and negative chirality Weyl points \cite{zyuzin-wsm, Balents2012, Balents2011}.

CS arrangements of Weyl points arise from Dirac points when time-reversal symmetry is broken. Experimental $\mathcal{T}$-breaking techniques include introducing an external magnetic field \cite{chiral-anomaly-factory}, doping with magnetic atoms \cite{eubaagbi-wan, eubaagbi-2}, or creating a layered heterostructure consisting of the Dirac semimetal and a magnetic insulator \cite{topological-lego}. We have computed the AHC and ANC for example systems in each of these three cases to demonstrate both their promise and unique challenges; details can be found in the supplementary material \cite{supplement}.

There are a number of general limitations to these approaches. 
Firstly, Dirac semimetals tend to have a low density of states at $E_F$, limiting their transport properties. Secondly, introducing magnetism to these systems causes a relatively small separation of the Weyl points in both energy and momentum space. While a small energy separation can be favorable for tuning the AHC and ANC, their magnitudes are proportional to the momentum-space Weyl separation. In order to achieve a large ANC, either the bands generating the Weyl points must be quite flat, resulting in a large density of states and momentum-space separation, or the overall number of Weyl pairs must be very large in order to compensate for the small contribution of each individual pair. 

When a narrow $d$ or $f$ electron band is partially occupied in a magnetic material, it can lead to very large numbers of Weyl points within a narrow energy range. In the next section, we present a magnetic material with a partially filled $d$-electron band hosting an arrangement of many Weyl points leading to a CS AHC and a large ANC, which we show can be further enhanced by manipulating the Weyl positions through strain.

\section{IV. Enhancing the ANC.}

Heusler materials have recently emerged as multi-functional platforms realizing a number of exotic effects, including a variety of topological features as well as large anomalous Hall and Nernst signals \cite{noky-heusler, giantANE-co2mnga}. When the symmetries of their cubic space group are broken via doping, strain, or magnetism, Heusler compounds can exhibit Dirac points, Weyl points, and nodal lines.

Ref. \cite{noky-heusler} scanned a database of Heusler compounds, identifying eight materials with relatively large AHC/ANC signals near the Fermi level stemming predominantly from Weyl points. We examined these eight materials for CS Weyl point configurations.  

\begin{figure}[ht]
    \includegraphics[width=1.0\columnwidth]{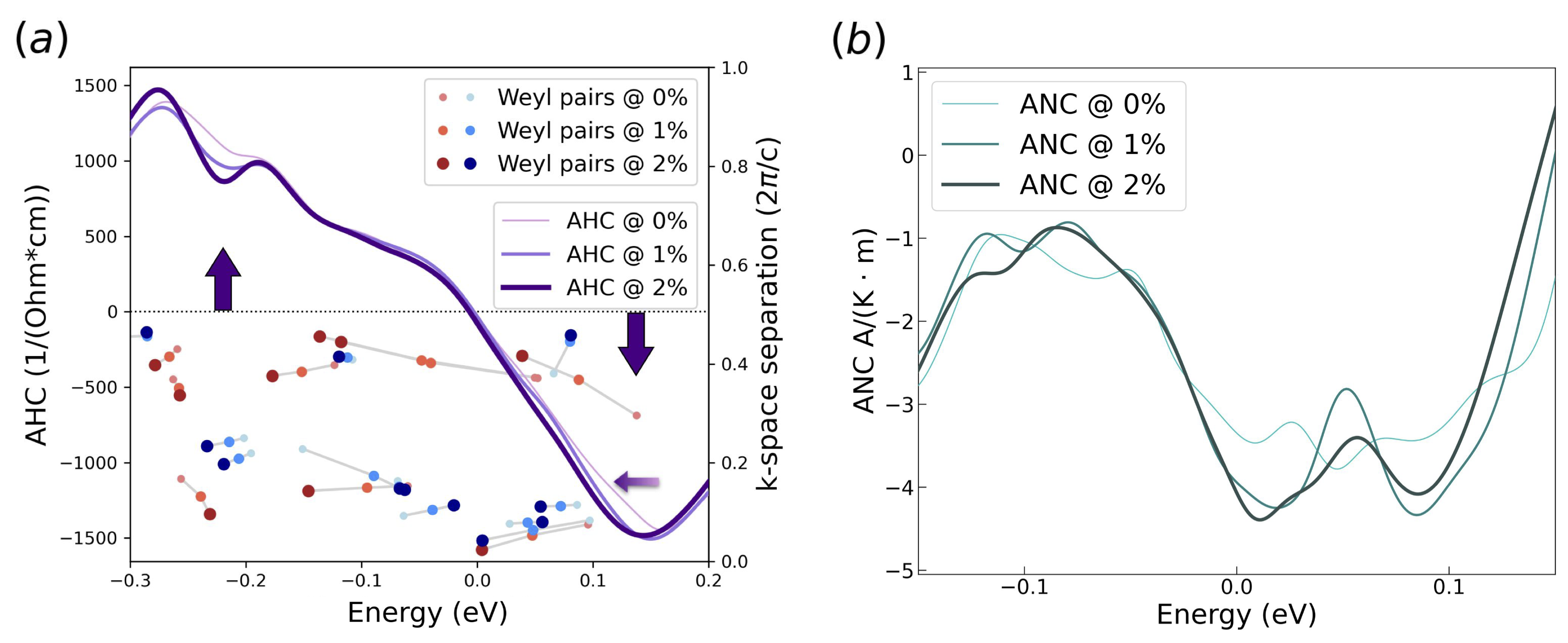}
    \caption{Computed (a) AHC overlaid with the pairs of Weyl points for different strains of the lattice and (b) ANC for each strain case. Both the AHC and ANC are calculated at 100 K. Note the peak-and-dip AHC signal around the Fermi level, characteristic of the CS feature, as in Fig. \ref{cartoon}(b), as well as the shift in the positive-energy peak with increasing strain.}
    \label{rh2nisi_bands}
\end{figure}

In order to identify CS features, density functional theory calculations were performed within the Vienna Ab-initio Simulation Package (VASP) using projector augmented wave pseudopotentials \cite{vasp1,vasp2,vasp3,vasp4}, and Perdew-Burke-Ernzerhof (PBE) functionals \cite{pbe}. Structural relaxation and subsequent electronic structure calculations used a $20 \times 20 \times 20$ $k$-point grid, with the magnetization constrained to the easy axis in the $\langle 001 \rangle$ crystallographic direction, and with spin-orbit coupling included. A plane-wave energy cutoff of 350 eV and convergence tolerance of $10^{-10}$ eV were used. The Wannierization step was carried out on the same $k$-point grid with the Wannier90 code \cite{w90,w90update}, using a convergence tolerance of $10^{-10}$ for disentanglement and $10^{-11}$ for Wannierization. The locations and chiralities of Weyl points were determined using the WannierTools package \cite{wanniertools}, which was also used to compute the AHC. A custom module was added to WannierTools to compute the ANC from the BC $\Omega_{ij}(\bm{k})$ using
\begin{align}
	\alpha_{ij} &= \frac{k_B e}{\hbar} \int \frac{d\bm{k}}{(2\pi)^{3}} \Omega_{ij}(\bm{k})\nonumber\\
	&\left[\ln(1 + e^{-\beta(\epsilon-\mu)}) +  \beta(\epsilon-\mu) f_{\text{FD}}(\bm{k}) \right],
	\label{anc_std}
\end{align}
where $\beta = 1/(k_B T)$. 

Our calculations confirm Rh$_2$NiSi as a magnetic ($\mu_\text{tot} \sim 0.99 \mu_B$) Weyl semimetal and identify a CS AHC, originating from a large number of Weyl points near the Fermi level.
The CS configuration for this material is optimal at temperatures far above the magnetic transition, but the ANC at lower temperatures can be enhanced by tuning the Weyl positions.

We begin with by discussing the transport properties of Rh$_2$NiSi in the absence of any external manipulation. The anomalous Hall spectrum of Rh$_2$NiSi, shown in Figure \ref{rh2nisi_bands}, shows that AHC is nearly zero at the Fermi level, framed on either side by a positive ($\sim 1500$ Ohm$^{-1}$ cm$^{-1}$) and negative ($\sim -1500$ Ohm$^{-1}$ cm$^{-1}$) peaks, in a CS configuration. 

While several techniques exist for manipulating the locations of Weyl points in materials \cite{weyl-mainip1, weyl-mainip2}, one of the simplest ways to perturb the electronic structure of a material is by applying strain. In order to mimic plausible experimental conditions, two epitaxial strains were applied, expanding the lattice by 1\% and 2\% respectively.


The locations of BC sources and sinks were computed within 500 meV of the Fermi level for each strain case,
and grouped into symmetry-equivalent sets of Weyl points according to the I4/mmm symmetry of the strained crystal. 
The positions of these Weyl points 
are provided in Table S1 \cite{supplement} and are reproduced in Figure \ref{rh2nisi_bands}b. We note that the large number of identified Weyl points is due to the metallic nature of the system, which leads to many band crossings near the Fermi energy.


The computed ANC for each strain is shown in Figure \ref{rh2nisi_bands}c. Under increasing strain the ANC at the Fermi energy increases from 3.35 A/(K m) to 4.07 A/(K m), while the peak ANC, just above $E_F$, similarly increases from 3.78 A/(K m) to 4.40 A/(K m).  
The increase in ANC can be understood qualitatively using the Mott relation. As strain is increased, the AHC peak above $E_F$ shifts towards the Fermi level (Fig. \ref{rh2nisi_bands}b), while the AHC peak below $E_F$ remains at the same energy, leading to a steeper gradient and thus an enhanced ANC. Another way to see this is that decreasing the gap $\Delta E$ between the peaks, pushes the system closer to maximal ANC (Eq. \ref{bound}), though it is still far from $\Delta E_\text{opt} = 26.6$meV which is optimal at 100K.





\begin{figure}[ht]
    \includegraphics[width=1.0\columnwidth]{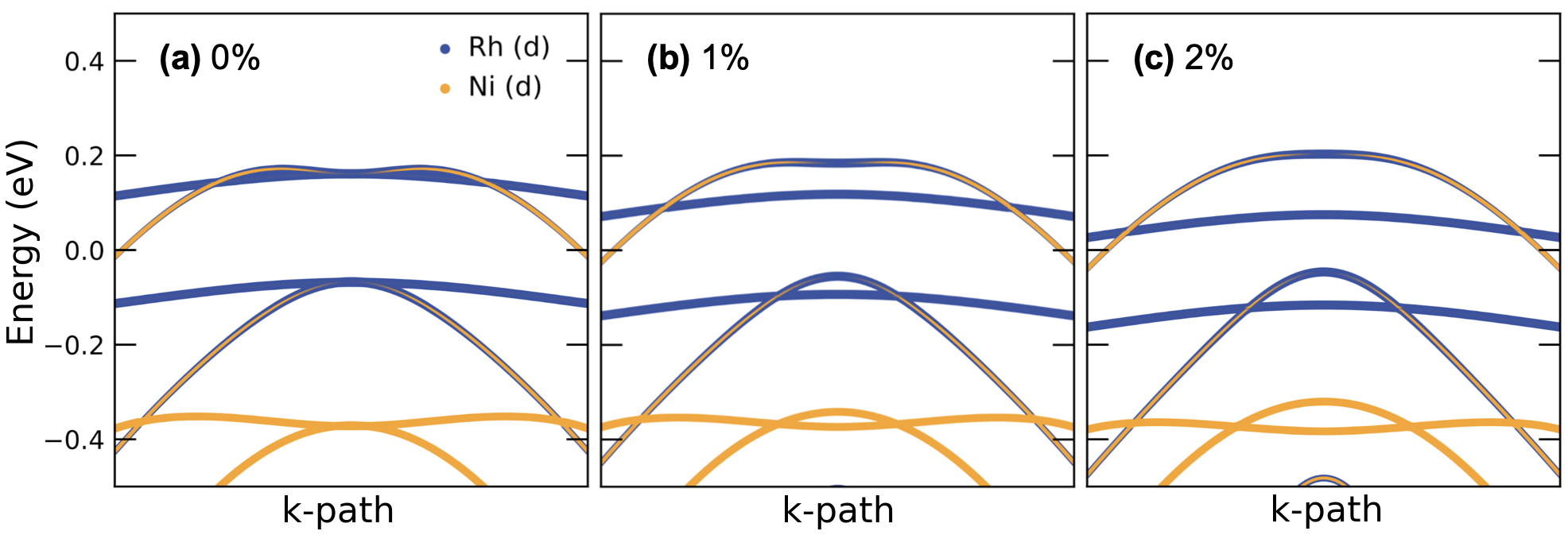}
    \caption{
        Projected band structure plot along the $\bm{k}$-path
        $\sum_i \bm{k}^i_\text{Weyl}\pm\alpha \hat{k}_z$ 
        near selected Weyl pairs of Rh$_2$NiSi. The most strain-sensitive Rh bands have mostly $d_{x^2-y^2}$ character.}
    \label{rh2nisi_weyls}
\end{figure}

The mechanism behind the motion of the AHC peaks can be understood within the context of the crystal field splitting and bandwidth energy parameters of the electronic structure of Rh$_2$NiSi. The cubic symmetry splits the Rh-$d$ band into $e_g$, and lower-lying $t_{2g}$ manifolds, which are occupied by the seven $d$-electrons of Rh, filling the $t_{2g}$ bands completely and the $e_g$ partially occupied. This single unpaired $e_g$ electron is responsible for the Rh local moment of $\sim 1 \mu_B$. 
At the intersection of Rh $e_g$ bands with Ni bands, the system forms regions of high BC, giving rise to the AHC peak at E=0.1 eV.
When the lattice is expanded in the $xy$-plane, the overlap between adjacent Rh-$d_{x^2-y^2}$ orbitals decreases, leading to a smaller hopping parameter and decreased crystal field interactions. As a result, the $e_g$ bands are narrowed and shifted downwards, moving the BC sources they form lower in energy.


This process suggests that the strain-dependent behavior of the CS arrangement in Rh$_2$NiSi is not simply serendipitous, but is in fact a consequence of the collective behavior of the Weyl points formed by the narrow Rh-$e_g$ bands. Most surprisingly, the considerable $\sim 20 \%$ increase in the ANC arises from comparatively small changes in the Weyl points.

\section{V Conclusion.}

The CS feature in the AHC presented here has been shown to amplify the anomalous Nernst conductivity. 
Furthermore, this feature allows for the tuning of the ANC to be maximal at a specific temperature, since the contributing Weyl points can be in different sets of bands. Such an arrangement of Weyls can be created by introducing magnetism into a material with Dirac points, either by external fields, doping, or interfacing with a magnet. It was also shown how in a material with a CS feature in the AHC and a large ANC, strain could be used to tune the locations of the Weyl points to bring the CS peak separation closer to the optimal $\Delta E_\text{opt}$ in order to further enhance the ANC. At room temperature, the condition $\Delta E_\text{opt} = 3.087 k_B T$ lies in an energy scale typical to the electronic structure of many materials. The challenge for future research is in engineering a band structure to produce a CS AHC distribution without extraneous peaks.

The constraint of Weyls leading to a CS AHC can be taken in combination with existing rules of thumb for materials with large ANC, including flat bands and strong spin orbit coupling. These rules inform future searches for improved thermoelectrics; for instance heterostructures like the proposed one with PdTe$_2$ and VI$_2$, can be scanned for such features near the Fermi energy. Another potential source of such materials are lanthanide compounds. While some lanthanide materials with Weyl points have been reported \cite{ybmnbi2, ceru4sn6, ceb6, ce113}, no concerted effort has been made to exhaustively search this space for topological Weyl materials. The narrow $4f$ bands of lanthanides have the potential to host large numbers of Weyl points closely spaced in energy, making them ideally suited for realizing configurations of Weyl points that lead to CS features which enhance the ANC.

\section{Acknowledgments}
This work was supported by the Office of Science, Office of Fusion Energy Sciences, of the U.S. Department of Energy, under Contract No. DE-AC02-05CH11231.  LZT and VI were also supported by the Molecular Foundry, a DOE Office of Science User Facility supported by the Office of Science of the U.S. Department of Energy under Contract No. DE-AC02-05CH11231. EB was supported by the National Science Foundation Graduate Research Fellowship under Grant No. 1752814.  This research used resources of the National Energy Research Scientific Computing Center, a DOE Office of Science User Facility supported by the Office of Science of the U.S. Department of Energy under Contract No. DE-AC02-05CH11231.

\vfill\null

\bibliography{references}

\end{document}